\documentclass[prb,twocolumn]{revtex4}

\usepackage{graphicx}
\def\CHR{Cr$_2$O$_3$}

\usepackage{amsmath}

\begin{document}

\title{Magnetism of chromia}

\author{Siqi Shi}

\altaffiliation[Present address: ]{Department of Physics, Center for
Optoelectronics Materials and Devices, Zhejiang Sci-Tech University, Xiasha
College Park, Hangzhou 310018, China.} \affiliation{Department of Physics and
Astronomy and Nebraska Center for Materials and Nanoscience, University of
Nebraska--Lincoln, Lincoln, Nebraska 68588, USA}

\author{A. L. Wysocki}

\affiliation{Department of Physics and Astronomy and Nebraska Center for
Materials and Nanoscience, University of Nebraska--Lincoln, Lincoln, Nebraska
68588, USA}

\author{K. D. Belashchenko}

\affiliation{Department of Physics and Astronomy and Nebraska Center for
Materials and Nanoscience, University of Nebraska--Lincoln, Lincoln, Nebraska
68588, USA}

\date{\today}

\begin{abstract}
The electronic structure and magnetism of chromia (corundum-type \CHR) are
studied using full-potential first-principles calculations. The electronic
correlations are included within the LSDA+U method. The energies of different
magnetic configurations are very well fitted by the Heisenberg Hamiltonian with
strong exchange interaction with two nearest neighbors and additional weak
interaction up to the fifth neighbor shell. These energies are insensitive to
the position of the oxygen states, indicating that magnetism in \CHR\ is
dominated by direct exchange. The N\'eel temperature is calculated using the
pair-cluster approximation for localized quantum spins of magnitude 3/2. Very
good agreement with experiment is found for all properties including the
equilibrium volume, spectral density, local magnetic moment, band gap, and the
N\'eel temperature for the values of $U$ and $J$ that are close to those
obtained within the constrained occupation method. The band gap is of the
Mott-Hubbard type.
\end{abstract}

\maketitle

Corundum-type \CHR\ is one of the antiferromagnetic transition-metal oxides
which present a significant challenge for electronic band theory due to the
correlated character of the partially filled, spin-polarized $3d$ shell. It
also has numerous applications in electronic devices, fuel cell electrodes, gas
sensors, heterogeneous catalysis, and thermal barrier coatings. Surface
properties of \CHR\ are of particular interest. The nominally polar, but
compensated (0001) surface exhibits structural phase transitions which are
poorly understood,\cite{Bender,Gloege} as well as an uncompensated surface
moment \cite{Binek} which may be useful in spintronic applications. It is
therefore very desirable to establish whether electronic correlations can be
reliably included in first-principles calculations in a way that would
accurately predict structural, electronic, and magnetic properties.

As expected for a transition-metal oxide, conventional density-functional
theory (DFT) studies of bulk \CHR\ \cite{Catti, Dobin, Wolter} have shown that
local spin density approximation (LSDA) or the generalized-gradient
approximation (GGA) for the exchange-correlation potential are unable to
reproduce the electronic and magnetic properties of bulk \CHR. Rohrbach
\emph{et al.} \cite{Rohrbach} performed a GGA+U calculation for \CHR\ using the
simplified (spherically averaged) $U-J$ correction \cite{Dudarev} and obtained
more reasonable results for the band structure. However, this approach is
inaccurate for structural and magnetic properties. First, as is typical for
transision-metal compounds, both GGA and the LSDA+U corrections reduce the LSDA
overbinding problem in \CHR\ and increase the equilibrium volume which becomes
close to the experimental value. However, the use of \emph{both} GGA and
Hubbard $U$ in the GGA+U method \cite{Rohrbach} results in an overcorrection,
so that the equilibrium volume becomes 7\% too large. Also, the magnetic
energies found by Rohrbach \emph{et al.} within this method are incompatible
with the experimental N\'eel temperature of about 308~K; they are too small
roughly by a factor of 5. Mosey \emph{et al.}\cite{Mosey} calculated the $U$
and $J$ parameters using the unrestricted Hartree-Fock method and found
$U-J=7.7$~eV, which is quite obviously too large for Cr, as is typical for
Hartree-Fock-like methods. They also studied the structural and electronic
properties of \CHR\ using the spherically averaged LSDA+U implementation
\cite{Dudarev} and found that the structural properties and the band gap come
out best at $U-J=4$~eV. For this value of $U-J$ the magnetic moment is somewhat
reduced from the ``ideal'' (ionic) value of 3 $\mu_B$ due to hybridization with
oxygen, which is also in agreement with experimental data.\cite{Note-LM}

In this paper we analyze the properties of \CHR\ in more detail with a
particular emphasis on its magnetism. We found that the LSDA+U method can
provide very good agreement with experiment for structural, spectral, and
magnetic properties with the same values of $U$ and $J$ which are close to
those found from the constrained occupation method within DFT.

The first-principles calculations were carried out using the projected
augmented wave (PAW) method \cite{Blochl, Kressej} implemented within the
Vienna \emph{ab initio} simulation package (VASP). \cite{Kresse, Kressef} The
$2s$ states on O were treated as valence states. We used the rhombohedral
primitive cell for the corundum structure in all calculations except those
involving complicated magnetic configurations (see below). The Monkhorst-Pack
scheme \cite{Monkhorst} based on the $4\times4\times4$ $k$-point grid was
employed for the Brillouin zone integrals, which were calculated using the
tetrahedron method with Bl\"ochl corrections. The plane-wave energy cutoff was
520 eV. These parameters ensured the total energy convergence to 2 meV/atom.
Densities of states (DOS) were calculated using the $8\times8\times8$ $k$-point
grid.

We employ the LSDA+U method in its full spherically symmetric form.
\cite{Liechtenstein} This extension is important for \CHR\ where correct
representation of crystal field and exchange splittings within the partially
filled $3d$ shell is critical. Surface energetics are even more sensitive to
the correct treatment of these unfilled shells; in fact, we found that the
errors introduced by the spherically averaged LSDA+U ansatz are intolerable for
the \CHR(0001) surface.\cite{Shi} The double-counting term is taken in the
fully localized limit. \cite{Liechtenstein,Petukhov}

Reasonable values of $U$ and $J$ can usually be obtained within DFT using the
constrained occupation method. \cite{Gunnarsson} These values are better suited
for use in LSDA+U calculations compared to the Hartree-Fock values, because
they include the screening of the $3d$ shell by the remaining electrons. We
calculated $U$ and $J$ in this way using the full-potential linear augmented
plane-wave (FLAPW) method implemented in the FLEUR package. \cite{FLEUR} In
these calculations the GGA approximation was used. We took the rhombohedral
primitive cell of \CHR\ containing four Cr atoms and set all the structural
parameters to their experimental values.\cite{Finger} The $3d$ electrons on one
or two (most distant) Cr atoms were formally treated as open core shells, i.e.
an integer occupation of these orbitals (for each spin projection) was
enforced, and their hybridization with all other electrons was turned off. The
$U$ and $J$ parameters are then found by comparing the LSDA total energies for
different charge and spin occupations of the $3d$ orbital(s) with their
Hartree-Fock expressions (the latter are equal to the ``double-counting'' terms
in LSDA+U).

The constrained occupation method is somewhat ambiguous because the $U$
parameter depends on the charge state of the ion \cite{Solovyev}. (On the
contrary, the $J$ parameter is usually very well defined; we found this to be
the case for \CHR\ as well.) Although the formal charge state of the chromium
ion is Cr$^{3+}$, we find $U$ and $J$ with respect to the $3d^4$ state. The
reason is that the screening properties of the valence electrons depend
primarily on the charge density distribution in the crystal, which is typically
very close to the superposition of atomic charge densities. Indeed, the formal
occupancy of the Cr $3d$ orbital within the 2.5 a.u. muffin tin sphere is about
4.2 in FLAPW. Specifically, the exchange parameter $J=0.58$~eV was found by
treating $3d$ electrons on one Cr atom as core and considering the energy
difference between the $3d^2_\uparrow3d^2_\downarrow$,
$3d^3_\uparrow3d^1_\downarrow$ and $3d^4_\uparrow3d^0_\downarrow$
configurations. The Hubbard parameter $U=3.3$~eV was found by treating the $3d$
shells on the two most distant Cr atoms as open cores and considering the
energy difference between the $3d_1^43d_2^4$ and $3d_1^53d_2^3$ configurations,
where 1 and 2 refer to the two different sites (the contribution of $J$ to the
energy differences was subtracted). Since the total number of electrons in the
cell is the same for both configurations, there is no need to include the Fermi
level correction.

\begin{figure}
\includegraphics*[width=0.45\textwidth]{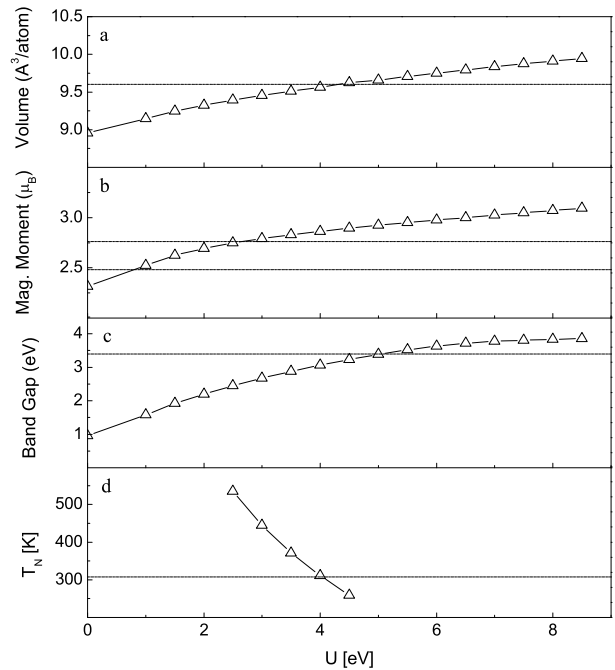}
\caption{Atomic volume (a), magnetic moment (b), band gap (c) and N\'eel
temperature (d) as functions of the Coulomb $U$ parameter for the
antiferromagnetic $+-+-$ state. The value of $J$ is fixed at 0.58~eV. The
horizontal lines denote experimental values; those in panel (b) are from both
Refs.\ \onlinecite{Brown,Corliss}.} \label{propU}
\end{figure}

In the following calculations we fixed $J$ at its calculated value of 0.58 eV
and varied $U$. The electronic structure is calculated in the ground
antiferromagnetic state with relaxed structural parameters (commonly denoted as
$+-+-$ in accordance with the ordering of Cr spins along the $z$ axis of the
rhombohedral cell). The dependence of the equilibrium volume, magnetic moment,
and band gap on $U$ is shown in Fig.\ \ref{propU}. We see that the volume and
the band gap agree quite well with experimental data at $U=4$ eV. Other
structural parameters also agree with experiment. The calculated angle between
the rhombohedral unit vectors is 55.11$^\circ$ compared to the measured
\cite{Finger} angle of 55.13$^\circ$. The shortest distance between Cr atoms
along the [111] axis is 2.646 \AA\ \emph{vs} the measured 2.650 \AA.

The magnetic moment at $U=4.0$ eV is 2.86 $\mu_B$, i.e. it is somewhat reduced
compared to the ``ideal'' ionic value of 3 $\mu_B$ corresponding to a fully
localized spin 3/2. Experimentally, the most recent neutron polarimetry
measurement gives the sublattice magnetization of 2.48 $\mu_B$ \cite{Brown},
which is notable lower compared to an older estimate of 2.76 $\mu_B$.
\cite{Corliss} The magnetic moment is smaller than 3~$\mu_B$ due to two
effects: (1) hybridization with oxygen, which is included in our calculation,
and (2) the quantum ``zero-point spin deviation,'' which is absent in DFT. The
zero-point deviation in \CHR\ was estimated \cite{Samuelsen} to be about 8\%,
which amounts to 0.24 $\mu_B$. Thus, keeping in mind the uncertainties related
to the \emph{definition}, measurement, and calculation of the magnetic moment,
its calculated value at $U\approx4.0$ eV is completely reasonable. Of course,
the poorly defined reduction from 3 $\mu_B$ can not be used as an indicator of
the quality of agreement with experiment. We also note that the local magnetic
moment depends very weakly on the magnetic configuration of \CHR. In the
ferromagnetic state the local moment within the muffin-tin sphere is 2.94
$\mu_b$, while the \emph{magnetization} is exactly equal to 3 $\mu_B$ per Cr
site, as expected. The orbital moment in the calculation with spin-orbit
coupling is small (about 0.04 $\mu_B$) and antiparallel to the spin moment, in
agreement with the experimental\cite{Foner} $g$-factor of 1.97 and with the
general rule for atomic shells that are less than half filled.

\begin{figure}
\includegraphics*[width=0.45\textwidth]{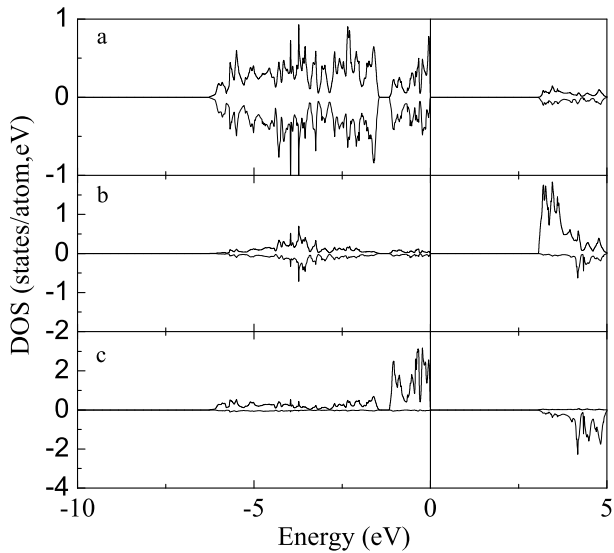}
\caption{Orbital projected DOS for the antiferromagnetic $+-+-$ ground state of
bulk \CHR, calculated within the LSDA+U with $U=4.0$~eV, $J=0.58$~eV. (a) O-2p,
(b) Cr-$e_{g}$ and (c) Cr-$t_{2g}$. Majority and minority-spin DOS are plotted
with different signs.} \label{dos}
\end{figure}

The band gap of 3.07 eV is somewhat smaller than the experimental value of 3.4
eV, but greater than that found in Ref.\ \onlinecite{Rohrbach} using GGA+U with
$U-J=4$~eV. Underestimation of the addition energy is a common feature of the
LSDA+U method, which is well known, for example, for Gd and other $4f$
elements. Further, the density of states shown in Fig.\ \ref{dos} is in
excellent agreement with X-ray photoemission data \cite{Zimmermann,Uozumi}.
Namely, the sharp and narrow peak at low binding energies separated by a
(pseudo)gap from the rest of the valence band is very well reproduced. Fig.\
\ref{dos} also shows the partial DOS decomposition into O and Cr contributions
from $t_{2g}$ and $e_g$ states. (The $t_{2g}$ and $e_g$ subbands are well
defined because the ligand field of the Cr site is approximately octahedral.)
As seen from Fig.\ \ref{dos}, the peak at low binding energies corresponds to
the filled Cr $t_{2g}$ spin subband with some admixture of oxygen $p$ orbitals.
At least some of this admixture is fictitious, because the Cr $d$
orbitals extend into oxygen's projection spheres. (The same ambiguity is
involved in the definition of the Cr magnetic moment.) Since non-magnetic \CHR\
is metallic with the Fermi level lying inside the $t_{2g}$ subband which is
separated by a gap from the oxygen $p$ band, the insulating gap forms by the
Mott-Hubbard mechanism.

Thus, the available structural and spectral properties of \CHR\ are well
reproduced by the spherically symmetric LSDA+U method with $U\approx4.0$ eV and
$J=0.58$ eV. This value of $U$ is somewhat larger than that given by the
constrained occupation method, but the latter was found using FLAPW
calculations with a different muffin-tin radius.

We now focus on the magnetic energetics which provides another stringent test
of the validity of the LSDA+U method for \CHR. We calculated the total energies
of 12 different magnetic configurations including the ferromagnetic state,
three simple antiferromagnetic orderings ($+-+-$, $++--$, and $+--+$), and
eight additional, arbitrarily chosen spin configurations in the hexagonal unit
cell of the corundum structure which includes six formula units. The ground
antiferromagnetic state has $+-+-$ ordering in agreement with experiment; its
relaxed structure was fixed for other magnetic orderings. The calculated total
energies are well fitted by the conventional Heisenberg Hamiltonian with
exchange interaction in five coordination spheres. The exchange parameters
fitted by the least squares method for different values of $U$ are listed in
Table \ref{jn}. The first, second and fifth nearest-neighbor interactions are
antiferromagnetic, whereas the third and fourth ones are ferromagnetic. All
exchange parameters except the third one reinforce the ground state. Also, we
see that the first and second exchange parameters are significantly larger than
those for more distant neighbors; this behavior is natural for an insulator. In
\CHR\ the parameter $J_1$ corresponds to the short bond along the $z$ axis, and
$J_2$ to the nearest neighbor bond within the buckled Cr layer. Each Cr atom
has one $J_1$ neighbor and three $J_2$ neighbors. As the value of $U$ is
increased, all antiferromagnetic exchange parameters systematically decrease,
while the ferromagnetic parameters systematically increase.

\begin{table}
\caption {The exchange parameters $J_n$ fitted using the total energies of 12
magnetic configurations calculated for different values of $U$ with $J$ fixed
at 0.58 eV. Each pair of sites contributes $J_{ij}\mathbf{e}_i\mathbf{e}_j$ to
the total energy, where $\mathbf{e}_i$ is the unit vector parallel to the
direction of the local moment at site $i$. The last column $\Delta$ shows the
mean-square misfit in the fitting of the total energy. $J_n$ and $\Delta$
are given in meV.}
\begin{ruledtabular}
\begin{tabular}{ccccccc}
$U$ & $J_1$ & $J_2$ & $J_3$ & $J_4$ & $J_5$ & $\Delta$ \\
\hline
2.5   & 30.9  & 21.9  & -0.60  & -1.83  & 4.92 & 0.91 \\
3.0   & 23.9  & 17.3  & -1.26  & -2.36  & 3.72 & 0.62 \\
3.5   & 18.6  & 13.8  & -1.74  & -2.72  & 2.84 & 0.43 \\
4.0   & 14.6  & 11.1  & -2.11  & -2.96  & 2.16 & 0.30 \\
4.5   & 11.1  & 9.04   & -2.41  & -3.11  & 1.64 & 0.20 \\
\end{tabular}
\end{ruledtabular}
\label{jn}
\end{table}

Let us compare our results with those derived from the spin wave dispersion
measured by inelastic neutron scattering. \cite{Samuelsen} The exchange
parameters fitted to those dispersion curves are also dominated by
antiferromagnetic $J_1$ and $J_2$. However, more distant neighbors in that
fitting \cite{Samuelsen} are much less important. For example, $J_5/J_1\approx
1/40$ in Ref. \onlinecite{Samuelsen} \emph{vs} $1/7$ found here. Also, there is
a notable disagreement in the $J_2/J_1$ ratio: Ref. \onlinecite{Samuelsen}
found $J_2/J_1\approx 0.45$ versus 0.76 for $U=4$~eV in Table \ref{jn}. We have
verified that the anisotropy $J_2/J_1$ is almost unchanged if only the
lowest-lying spin configurations are included in the fitting.

We also considered the possibility that the exchange parameters may be affected
by lattice distortion below the N\'eel temperature due to magnetostructural
coupling. In order to study this possibility, we need to know how the exchange
parameters depend on the structural parameters, and how the latter change
between room temperature and liquid nitrogen temperature where the spin wave
spectrum was measured. We found that the values of $J_1$ and $J_2$ are very
sensitive to the lattice parameters; a 1\% increase or decrease in the first or
second neighbor bond length leads to a 25-50\% decrease or increase of the
corresponding exchange parameter (this was roughly established by varying the
Bravais lattice parameters of the rhombohedral cell and fitting the four simple
magnetic orderings to the Heisenberg model). Thus, the $J_2/J_1$ ratio is very
sensitive to the $c/a$ ratio (or any lattice distortion that changes the ratio
of the first and second bond lengths $R_2/R_1$).

To our knowledge, no experimental data are available on the thermal contraction
of \CHR\ below room temperature. In order to assess the degree of
magnetostructural coupling, we have fully relaxed the structure for all the 12
magnetic configurations that were included in the fitting of exchange
parameters. The magnetostructural coupling is, in general, quite appreciable;
the $R_2/R_1$ ratio varies between the minimum of 1.072 in the $+--+$
configuration (where the nearest-neighbor pairs are all parallel, and
next-nearest antiparallel) and the maximum of 1.103 in the $++--$ configuration
(where the situation is reversed). The overall trend, as expected in general
for antiferromagnetic coupling, is for each of these bonds to shorten when the
corresponding spins are antiparallel and lengthen when they are parallel.
Although this effect seems to be rather large, the actual $+-+-$ state has
\emph{both} first and second-neighbor pairs antiparallel, so that both should
lengthen somewhat in the paramagnetic state. Among our 12 configurations, 5
have the same ratio $N^{(i)}_P/N^{(i)}_{AP}$ of the number of parallel and
antiparallel pairs in the $i$-th coordination sphere for $i=1$ and $i=2$, i.e.
these 5 configurations approximately represent the change of structure as a
function of the antiferromagnetic order parameter. The $R_2/R_1$ ratio for
these 5 structures varies between 1.088 in the $+-+-$ state and 1.084 in the
ferromagnetic state. The 3 intermediate states, all of which have twice as many
parallel pairs than antiparallel ones, all have $R_2/R_1=1.087$. The bond
lengths themselves change by about 0.7\% between the $+-+-$ state and these
three intermediate states. This variation is obviously too small to explain the
disagreement with experiment in the $J_2/J_1$ ratio.

Thus, in spite of the overall agreement with experiment for most
properties, a discrepancy in the $J_2/J_1$ ratio remains. It is possible that
our Heisenberg model fitting is not fully applicable to small deviations from
the ground state, for which the linear response technique should be a better
fit. Further, spin waves are more sensitive to distant couplings compared to
the energy fits or thermodynamic properties, which makes the fits from spin
wave data and from the overall energetics statistically inequivalent. Finally,
many-body effects beyond LSDA+U may play a role in renormalizing the exchange
parameters. We also note that impurities and thermal spin disorder should be
more effective in destroying distant couplings compared to nearest-neighbor
ones; this may explain the larger role of couplings beyond $J_2$ in our fitting
compared to experimental spin wave results.

The mechanisms of exchange interaction in \CHR\ are not well understood.
Complicated crystal structure with the presence of many electronic orbitals of
different symmetry and many Cr-O-Cr links at different angles greatly
complicates the empirical analysis. Both superexchange and direct exchange
interactions have been invoked to explain the magnetic structure of \CHR.
\cite{Li,Iida,Goodenough,Osmond} Direct exchange interaction may be interpreted
in terms of hopping of electrons from one Cr ion to another across the
insulating gap; the energy denominator involved in this process is the
Mott-Hubbard splitting. The hopping can only be effective between the orbitals
that are able to hybridize. From the DOS decomposition shown in Fig.\ \ref{dos}
it is clear that the $t_{2g}$ orbitals very weakly hybridize with $e_g$
orbitals on the neighboring Cr atoms. Therefore, the contribution of $e_g$
orbitals to direct exchange can be neglected. The $t_{2g}$ subband is split off
by crystal field and exactly half-filled, therefore direct exchange should be
antiferromagnetic. The superexchange involves hopping between Cr and O ions;
the energy denominator involves the gap between the oxygen $p$ states and the
unoccupied Cr states.

In order to reveal the mechanism responsible for exchange interaction in \CHR,
we use the following trick. A fictitious external potential $V$ is
coupled to the oxygen $p$ orbitals, which adds the term
$E_V=V\mathop\mathrm{Tr}n^\sigma_{mm'}$ to the total energy, where the trace is
taken over orbital and spin indices, and $n^\sigma_{mm'}$ is the density matrix
of the oxygen $p$ states defined inside the muffin tin sphere of 1.2 a.u. This
density matrix is calculated using the standard LSDA+U machinery. When $V$ is
set to a negative value, the filled oxygen $p$ states are pushed down to lower
energies, which suppresses superexchange, but not direct exchange. Weak
hybridization between the filled $t_{2g}$ states and the oxygen $p$ states
pushes them apart at $V=0$. When the $p$ states are pushed down by increasing
$V$, this repulsion is reduced and the $t_{2g}$ states also move down, thereby
increasing the band gap. Since direct exchange is sensitive to the band gap,
for better comparison we compensate this increased band gap by reducing $U$ on
Cr. This is done in such a way that the distance between the center of mass of
the filled $t_{2g}$ band and the conduction band minimum (CBM) remains the same
as at $V=0$.

Starting from the state with $U=3.5$ eV, we added $V$ of $-12$ eV and $-24$ eV
and calculated the energies of four magnetic states in the rhombohedral
primitive cell of \CHR. These calculations were performed using the FLAPW
method;\cite{FLEUR} the results are listed in Table \ref{ShiftO}. (Those for
$V=0$ are within 10\% of VASP results.) The self-consistent downward shift of
the $p$ states is much smaller than $Vn_{mm}$ because $V$ is strongly screened.
The distance $\Delta$ from the oxygen $p$ band maximum to CBM is also listed in
the table.

\begin{table}
\caption {Energies of three simple magnetic configurations relative to the
ground $+-+-$ state (in meV per formula unit) as a function of the fictitious
external potential $V$ applied to the oxygen $p$ orbitals (in eV). Reduced
values of $U$ on Cr compensate for the increased band gap (see text for
details). $\Delta$ is the distance from the oxygen band top to CBM in eV.}
\begin{ruledtabular}
\begin{tabular}{cccccc}
$V$ & $U$ & $\Delta$ & $++++$ & $++--$ & $+--+$ \\
\hline
  0 & 3.5 & 4.5 & 130 & 124 & 66 \\
-12 & 3.5 & 5.8 & 115 & 102 & 59 \\
-12 & 2.5 & 5.6 & 154 & 128 & 74\\
-24 & 3.5 & 7.9 & 102 & 80 & 51 \\
-24 & 2.15 & 7.7 & 145 & 110 & 71
\end{tabular}
\end{ruledtabular}
\label{ShiftO}
\end{table}

The results listed in Table \ref{ShiftO} clearly show that the magnetic
energies are insensitive to the position of the oxygen $p$ band. The reduction
of magnetic energies produced by adding negative $V$ at constant $U$ is due to
the fact that the $t_{2g}$ band gap increases due to dehybridization from
oxygen. Once $U$ is decreased to bring the band gap to its original value, the
magnetic energies are essentially unchanged compared to their values at $V=0$;
in fact, they even increase somewhat. On the other hand, we've seen above that
the magnetic energies are very sensitive to the value of $U$ which is
responsible for the band gap. This behavior leads us to a striking conclusion
that, contrary to the common belief, superexchange plays no role in magnetism
of \CHR. Antiferromagnetism is due exclusively to direct exchange, which, as
mentioned above, is antiferromagnetic because the magnetically active $t_{2g}$
subband is half-filled. It is likely that superexchange in \CHR\ is highly
ineffective because the Cr-O-Cr angles are close to 90$^\circ$, while the
overlap between O states and Cr $t_{2g}$ states is small. On the other hand,
the overlaps between $t_{2g}$ states on neighboring Cr atoms are quite large;
the $t_{2g}$ bandwidth in the ferromagnetic state at $V=-24$~eV is
approximately 1.5 eV.

We now calculate the N\'eel temperature $T_N$. We saw above that the local
moments on Cr atoms are well localized, and the energies of spin configurations
are well represented by the Heisenberg spin Hamiltonian. We therefore adopt the
quantum Heisenberg model for localized spins of magnitude $3/2$ with the
exchange parameters listed in Table \ref{jn}. Since each spin is strongly
coupled only to four neighbors (one with $J_1$ and three with $J_2$), the
mean-field approximation can not be reliably used. However, the
antiferromagnetic interaction is not frustrated. The important spin
correlations should be generated by the dominant exchange interaction with four
nearest neighbors. The network of bonds corresponding to $J_1$ and $J_2$ is
very weakly connected; the shortest closed path on this network is 6 bonds
long. Therefore, it is sufficient to capture the pairwise spin correlations. In
this situation, the pair cluster approximation appears to be an obvious choice.
This approach is a special case of the cluster variation method when the set of
clusters includes only pairs of sites. Here we follow the formulation of Ref.\
\onlinecite{Vaks} which can be directly applied to our case. The details of
this technique are described in the Appendix.

The calculated $T_N$ is shown in Fig.\ \ref{propU} as a function of $U$. We see
that the best agreement with experiment for $T_N$ is obtained at the same value
of $U\approx4$ eV as for the structural and spectral properties explored above.
This overall agreement is a strong indication that the essential details of the
electronic structure of \CHR\ are very well captured by the LSDA+U
approximation. Physically, this success of LSDA+U is explained by the presence
of fully filled and empty subbands which are strongly split by crystal and
exchange fields; LSDA+U usually reproduces such closed atomic-like subshells
very well.

In conclusion, we found that the spherically symmetric LSDA+U method provides a
very good description of structural, spectral, and magnetic properties of
chromia with $J=0.58$ eV found from the constrained occupation method and
$U\approx4$ eV, which is also close to the calculated value. We found that the
magnetic energies are well represented by a Heisenberg model with strong
exchange interaction with nearest neighbors both in the plane and along the $z$
axis and much weaker interaction with more distant neighbors. The artificial
downward shift of the filled oxygen $p$ states has almost no effect on the
magnetic energies, which proves that direct exchange is the dominant mechanism
of magnetic interaction.

This work was supported by NRC/NRI supplement to NSF-MRSEC and by the Nebraska
Research Initiative. K. B. is a Cottrell Scholar of Research Corporation.

\appendix*

\section{}

Here we describe the application of the pair cluster approximation to \CHR\
following a similar formalism of Ref.\ \onlinecite{Vaks}. The energy of a
quantum Heisenberg magnet (per unit cell) can be written as
\begin{equation}
E=-\frac12\sum_{i,j}m_in_{ij}J_{ij}\langle\hat{\mathbf{S}}_i\cdot\hat{\mathbf{S}}_j\rangle-\sum_{i}m_iB_i\langle
\hat S_i^z\rangle, \label{energy}
\end{equation}
where the summations are over the inequivalent sites in the unit cell,
$\hat{\mathbf{S}}_i$ are quantum spin operators, $m_i$ is the number of sites
of type $i$ in the unit cell, $n_{ij}$ the number of neighbors of site $i$ that
are of type $j$, and $B_i$ the external magnetic field applied to site type
$i$. All the thermodynamic properties can be obtained from the free energy
which may be calculated by integrating the Gibbs-Helmholtz relation:
\begin{equation}
F=\frac{1}{\beta}\int_0^{\beta}E(\beta')d\beta', \label{freeenergy}
\end{equation}

To proceed we need to find the expectation values appearing in Eq.
(\ref{energy}). In the pair-cluster approximation they are calculated by
introducing one- and two-site clusters with the following cluster Hamiltonians:
\begin{eqnarray}
\hat H_1^i &=& -h_i\hat S_i^z\nonumber\\
\hat H_2^{ij} &=&
-J_{ij}\hat{\mathbf{S}}_i\cdot\hat{\mathbf{S}}_j-h_i^{(j)}\hat
S_i^z-h_j^{(i)}\hat S_j^z, \label{clusters}
\end{eqnarray}
where $h_i=B_i+\phi_i$ is the one-site ``cluster field,''
$h_i^{(j)}=h_i-\phi_i^{(j)}$ is the cluster field at site $i$ for the pair
cluster ($i$,$j$). The one-site and two-site cluster fields are related through
$\phi_i=\sum_{j}n_{ij}\phi_i^{(j)}$. The quantities $\phi_i^{(j)}$ are treated
as variational parameters and found by minimizing the free energy, i.e.
requesting that $\partial F/\partial \phi_i^{(j)}=0$. It can be shown that this
variational condition ensures that the expectation value $\langle \hat
S_i^z\rangle$ is the same in all one-site and all two-site clusters containing
site $i$. The expectation values
$\langle\hat{\mathbf{S}}_i\cdot\hat{\mathbf{S}}_j\rangle$ are calculated from
the pair cluster ($i$,$j$). Performing the integration in (\ref{freeenergy}),
we find
\begin{equation}
F=-\frac{1}{2\beta}\sum_{ij}m_in_{ij}\ln Z_2^{ij}
+\frac{1}{\beta}\sum_im_i(n_i-1)\ln Z_1^i, \label{F}
\end{equation}
where $n_i=\sum_j n_{ij}$ is the total number of neighbors of site $i$, while
$Z_1^i=\mathop\mathrm{Tr}\exp(-\beta H_1^i)$ and
$Z_2^{ij}=\mathop\mathrm{Tr}\exp(-\beta H_2^{ij})$ are the one-site and
two-site cluster partition sums. Evaluation of $Z_1^i$ is trivial; to find
$Z_2^{ij}$ one first needs to diagonalize $H_2^{ij}$.

Our goal here is to find the transition temperature; therefore we may assume
that all $h_i^{(j)}$ are small. The free energy is developed in these
parameters, and then the parameters $\phi_i^{(j)}$ are found by requiring that
the variation of the free energy $F$ vanishes. (The resulting equations are too
cumbersome to be included here.)

For \CHR\ we assume the actual magnetic ordering $+-+-$. All Cr sites are
related by magnetic symmetry, which reduces the number of independent
variational parameters. The transition temperature is found by setting $S=3/2$
and searching for the pole of the magnetic susceptibility which is found from
the one-site cluster:
\begin{equation}
\chi_{ij}=\frac{\partial \langle S_i^z \rangle}{\partial
B_j}=\frac{S(S+1)}{3}\beta \frac{\partial h_i}{\partial B_j}. \label{susc}
\end{equation}

The resulting equation for the transition temperature has two solutions. The
greater solution is $T_N$, while the lower one is the fictitious ``anti-N\'eel
point'' below which $\langle \hat S_i^z \rangle=0$. The existence of an
anti-N\'eel point for antiferromagnets is a well-known shortcoming of the
pair-cluster approximation, \cite{Kasteleijn,Vaks} which fails at low
temperatures. However, in our case the anti-N\'eel point is much smaller than
$T_N$ which indicates that the pair cluster approximation has a wide range of
validity. Therefore, this method is expected to provide a very good
approximation for $T_N$.


\begin{thebibliography}{99}
\bibitem{Note-LM} Refs. \onlinecite{Rohrbach} and \onlinecite{Mosey}
concluded that there is a large disagreement with experiment in the magnetic
moment, but they used an incorrect experimental value of 3.8 $\mu_B$.
\end{thebibliography}
\end{document}